\documentclass[preprint,showpacs]{revtex4}
\usepackage{epsfig}
\usepackage{amsmath}
\usepackage{caption}

\begin{document}

\author{D. Froemberg}
\affiliation{Institut f\"ur Physik,
Humboldt-Universit\"at zu Berlin, Newtonstra\ss e 15, 12489 Berlin, Germany}
\author{H. Schmidt-Martens}
\affiliation{Institut f\"ur Physik,
Humboldt-Universit\"at zu Berlin, Newtonstra\ss e 15, 12489 Berlin, Germany}
\author{I.M. Sokolov}
\affiliation{Institut f\"ur Physik,
Humboldt-Universit\"at zu Berlin, Newtonstra\ss e 15, 12489 Berlin, Germany}
\email{e-mail: igor.sokolov@physik.hu-berlin.de}
\author{F. Sagu\'es}
\affiliation{Departament de
Qu\'imica F\'isica, Universitat de Barcelona, Mart\'i i Franqu\`es
1, E-08028, Barcelona, Spain}

\title{Front propagation in A+B $\rightarrow$ 2A reaction under subdiffusion}


\begin{abstract}
We consider an irreversible autocatalytic conversion reaction
$\mathrm{A} +  \mathrm{B} \rightarrow 2\mathrm{A}$ under subdiffusion
described by continuous time random walks. The reactants'
transformations take place independently on their  motion and are
described by constant rates. The analog of this reaction in the
case of normal diffusion is described by the Fisher-Kolmogorov-Petrovskii-Piskunov 
(FKPP) equation leading to the existence of a nonzero minimal front propagation
velocity which is really attained by the front in its stable motion.
We show that for subdiffusion this minimal propagation velocity is zero,
which suggests propagation failure. 
\end{abstract}

\maketitle

\section{Introduction}
The theory of reactions under subdiffusion had attracted recently
considerable attention both because of theoretical and mathematical
challenges posed by such problems, and also due to their growing
practical relevance for description of phenomena taking place in
porous media(as exemplified by geophysical structures) and in crowded
interior of living cells (e.g. Refs. \cite{koszt, goldingcox,
banks}). Several recent works were dedicated to the theoretical
description of Turing patterns and of fronts in such systems
\cite{horsthemke, TurHLW, AB0, YAD}. Thus Ref.\cite{YAD} concentrates
on the front behavior in the system which in the case of the normal
diffusion would be described by the
Fisher-Kolmogorov-Petrovskii-Piskounov (FKPP) equation and shows, that
under the description adopted, there exists a (minimal) stable
propagation velocity of such front, just like it is the fact under
normal diffusion. As we proceed to show, this is not always the
case. Considering the fully irreversible analog of the reaction
discussed in Ref.\cite{YAD} under conserved overall concentration we
show that the minimal propagation velocity is zero, which corresponds
essentially to propagation failure. The system we consider here
corresponds to the $\mathrm{A}+\mathrm{B} \rightarrow 2\mathrm{A}$
irreversible reaction, which in the case of the normal diffusion is
also described by the FKPP equation.

The FKPP equation \cite{FKPP}, proposed by R. A. Fisher in 1937
\cite{fisher} as a model for propagation of a favorable gene in a
population, corresponds to a mathematical description of the
(irreversible or reversible) reaction whose main stage is a
bimolecular autocatalytic conversion $\mathrm{A}+ \mathrm{B}
\rightarrow 2 \mathrm{A}$. Initially the whole system consists of
particles (individuals) of type B. The introduction of the
A-individuals into some bounded spacial domain (which is described by
an initial condition sharply concentrated in vicinity of the origin of
coordinates) leads to a propagation of a front of A into the
B-domain. Physically this corresponds to a front propagating into the
unstable state. Under normal diffusion, the process is described by a
partial differential equation
\[
\frac{\partial A}{\partial t} = D \frac{\partial^2 A}{\partial x^2} + kAB
\]
for an irreversible reaction, where the initial concentration
of B is assumed to be homogeneous and equal to $B(x,0)=B_0$ everywhere except for
the vicinity of the origin. In this case the overall concentration is conserved, due to
the local stoichiometry of the reaction which does not change the number of
particles. Here and in what follows we denote the nature of particles in reaction
equations by Roman letters, while the corresponding concentrations or particle numbers
are written in \textit{italic}.

Using the conservation law,
the corresponding reaction-diffusion equation can be rewritten as
\begin{equation}
\frac{\partial A}{\partial t} = D \frac{\partial^2 A}{\partial x^2} + kAB_0-kA^2.
\label{FKPPgen}
\end{equation}
The equation for the reversible reaction has the same form but different 
coefficients in front of the two last terms in the right hand side (e.g. \cite{RioDoe}).
The FKPP equation is the simplest model of front propagation into an unstable state
and serves as a paradigmatic model for many related phenomena.

The front velocity in the genuine FKPP equation is determined by that
of its leading edge, i.e. by the behavior of concentrations for $x
\rightarrow \infty$. Since the concentration of converted particles in
this leading edge is very small, the equation can be linearized, and
the possible velocity of the front is given by the analysis of the
\textit{linear} propagation problem \cite{panja, saarl}. For the FKPP
equation, Eq.(\ref{FKPPgen}) leads for $x \rightarrow \infty$ (and $B
\rightarrow B_0$) to
\begin{equation}
\frac{\partial A}{\partial t} = D \frac{\partial^2 A}{\partial x^2} + kB_0 A
\label{FKPPlin}
\end{equation}
(or by the corresponding equation for $\delta B = B_0 - B = A$)
having the exponential propagating solution $A \simeq
\exp(-x+vt)$ for all $v \geq 2\sqrt{DkB_0}$ where the condition on $v$
is imposed by the natural boundary condition $A =0$ for $x
\rightarrow \infty$ and by forbidding oscillatory behavior. 
Further analysis shows that the minimal
velocity $v = 2\sqrt{DkB_0}$ is exactly the one attained, which fact
is known as the marginal stability principle. The front in the FKPP
system is an example of the so-called ``pulled front'', as it is
``pulled'' into the unstable state by its leading edge, and its
propagation velocity does not depend on what happens in the interior
of the front where the conversion of the most particles takes place.

Modeling reaction phenomena in subdiffusive media requires some
preliminary assumptions on the nature of the transport
process. Parallel to Refs.\cite{SSS,sashso} we assume that the
subdiffusive motion on a mesoscopic scale is a consequence of trapping
of particles on a mesoscopic scale, due to e.g. the bottlenecks
connecting the voids in a porous system, while on a microscopic scale
within the pores the $\mathrm{A}+\mathrm{B} \rightarrow 2\mathrm{A}$
reaction takes place in a homogeneous solution and follows the mass
action law. The local conservation of particle concentration is an
inherent property of such systems which also holds on the mesoscopic
scale. The model adopted corresponds therefore to the systems which at
smaller scales consists of compartments in which the reaction follows
the usual kinetic laws, while the subdiffusive transport between the
compartments is described by continuous time random walks with a
distribution of sojourn times which may lack the first moment. As we
proceed to show, due to the coupling between the reaction and
transport term inherent for subdiffusion \cite{SSS,sashso}, the
behavior of the reaction front under subdiffusion and under the
conditions discussed above is vastly different from the one under
normal diffusion. The minimal propagation velocity of the front is
zero, which corresponds to propagation failure.  The preliminary
results of our numerical simulations confirm that the front's velocity
decays with time. These numerical results will be discussed in detail
elsewhere.

\section{The $\mathrm{A}+\mathrm{B} \rightarrow 2 \mathrm{A}$ reaction under subdiffusion}

\subsection{General considerations}
Following the same procedure as in \cite{AB0, SSS} one obtains the
equations for the particle concentrations as a consequence of the
conservation laws. Discretizing the system into compartments (sites)
numbered by the index $i$ we get for the mean numbers of B-particles
at site $i$
\begin{equation}
\dot{B}_i(t) = \frac{1}{2} j_{i-1}^{-}(t) + \frac{1}{2} j_{i+1}^{-}(t) - j_{i}^{-}(t) - k A_{i}(t) B_{i}(t). \label{fkppbalB} 
\end{equation}
Eq. \ref{fkppbalB} is a local balance equation for the number of
B-particles at site $i$ with $j_i^-(t)$ being loss fluxes of particles
B from the site $i$ at time $t$ given by
\begin {eqnarray}
j_{i}^{-}(t) &=& \psi (t) P_{B}(t,0) B_{i}(0) \nonumber\\ 
&& + 
\int_{0}^{t}\psi(t-t^{\prime}) P_{B}(t,t^{\prime}) \left[ \dot{B}_i(t^{\prime}) + j_{i}^{-}(t^{\prime}) + k A_{i}(t^{\prime}) B_{i}(t^{\prime}) \right] dt^{\prime} 
\label{fkppjB} \\
P_{B}(t,t^{\prime}) &=& \exp\left[ -k \int_{t^{\prime}}^{t}A_{i}(t^{\prime\prime}) dt^{\prime\prime} \right]. \label{fkppSPB}
\end {eqnarray}
 The
first term on the right hand side of Eq.(\ref{fkppjB}) for the
corresponding flux gives the contribution of those
$\mathrm{B}$-particles that were at $i$ from the very beginning and
survived until $t$. The second term describes the particles, that
arrived at $i$ at a time $t^{\prime}$ and did neither react nor
perform a jump until $t$.  Equation (\ref{fkppSPB}) gives the survival
probability of B-particles.  The $\mathrm{A}$-concentration is then
given by
\begin {eqnarray}
\dot{A}_i(t) &=& \frac{1}{2} f_{i-1}^{-}(t) + \frac{1}{2} f_{i+1}^{-}(t) 
- f_{i}^{-}(t) + k A_{i}(t) B_{i}(t) \label{fkppbalA} \\
f_{i}^{-}(t) &=& \psi (t) P_{A}(t,0) B_{i}(0) + \psi(t) A_{i}(0) \nonumber \\
& & + \int_{0}^{t}\psi(t-t^{\prime}) \left[ \dot{A}_i(t^{\prime}) + f_{i}^{-}(t^{\prime}) 
- k A_{i}(t^{\prime}) B_{i}(t^{\prime}) \right] dt^{\prime} \nonumber \\
& & +  \int_{0}^{t}\psi(t-t^{\prime}) P_{A}(t,t^{\prime}) \left[ \dot{B}_i(t^{\prime}) 
+ j_{i}^{-}(t^{\prime}) + k A_{i}(t^{\prime}) B_{i}(t^{\prime}) \right] dt^{\prime} 
\label{fkppjA} \\
P_{A}(t,t^{\prime}) &=& 1 - P_{B}(t,t^{\prime}) \label{fkppSPA}
\end {eqnarray}
The loss fluxes for A-particles are denoted through $f_i(t)$. The
first term on the right hand side of Eq.(\ref{fkppjA}) corresponds to the
particles that were at $i$ from the beginning and converted from
$\mathrm{B}$ to $\mathrm{A}$ until $t$. The second term represents the
$\mathrm{A}$-particles that were at $i$ from the very beginning. The
third and fourth term describe the particles that arrived at $i$ at a
time $t^{\prime}$ as $\mathrm{A}$-particles, or arrived as
$\mathrm{B}$-particles and reacted until $t$. The probability $P_{A}$
to gain new $\mathrm{A}$-particles arises from the conservation of the
total number of particles and the probability $P_{B}$ that the
$\mathrm{B}$-particles react. The $\mathrm A$-concentration depends on
the $\mathrm B$-concentrations at all previous times.

Equations (\ref{fkppbalB}--\ref{fkppSPA}) are consistent with
the conservation of the total number of particles in the reaction
$\mathrm{A}+\mathrm{B} \rightarrow 2 \mathrm{A}$. Let $C_i=A_i+B_i$ be
the total mean particle number at $i$ and
$g_{i}^{-}(t)=f_{i}^{-}(t)+j_{i}^{-}(t)$ the total loss flux of
particles at $i$. From Eqs. (\ref{fkppbalB}, \ref{fkppbalA}) we obtain
the balance equation:
\begin{equation}
\dot{C}_{i}(t) = g_{i}^{+}(t) - g_{i}^{-}(t) = \frac{1}{2} g_{i-1}^{-}(t)+\frac{1}{2} g_{i+1}^{-}(t) - g_{i}^{-}(t).
\end{equation}
From Eqs.(\ref{fkppjB}, \ref{fkppjA}) follows that
\begin{eqnarray}
g_{i}^{-}(t) &=&  \psi(t) [B_{i}(0) + A_{i}(0)] + 
\int_{0}^{t}\psi(t-t^{\prime}) \nonumber \\ 
& &\times \left[ \dot{B}_{i}(t^{\prime}) + 
\dot{A}_{i}(t^{\prime}) +  f_{i}^{-}(t^{\prime}) + 
j_{i}^{-}(t^{\prime}) \right] dt^{\prime} \nonumber \\ 
&=&  \psi(t) C_{i}(0) + \int_{0}^{t}\psi(t-t^{\prime}) 
\left[ \dot{C}_{i}(t^{\prime}) + g_{i}^{-}(t^{\prime}) \right] dt^{\prime}. \label{fkppges}
\end{eqnarray}
This equation can be solved by means of the Laplace transform,
\begin{equation}
\tilde{g}_{i}^{-}(u) = \frac{u\tilde \psi(u)}{1-\tilde \psi(u)}\tilde C_i(u),
\end{equation}
which yields a diffusion equation for $C_i(t)$ in the Markovian case 
$\psi(t) =\frac{1}{\tau}\exp{\left[ -\frac{t}{\tau}\right] }$, and a
subdiffusion equation in the non-Markovian case $\psi(t) \propto
t^{-1-\alpha}$, i.e. shows that the behavior of the total particle concentration is
diffusive or subdiffusive, respectively. Moreover, if we choose the
initial condition in a way that $C_{i}(t)=const$, the total number of
particles is also locally conserved, $C_i(t)=C_i(0)=C_0$.
This reduces the overall problem to the one
for only one species. Passing
to the continuous variables $x=ai$ we obtain:
\begin {eqnarray}
\dot{B}(x,t) &=& \frac{a^2}{2} \Delta j^{-}(x,t) - k (B_0 - B(x,t)) B(x,t) \nonumber \\
j^{-}(x,t) &=& \psi (t) P_{B}(x,t,0) B(x,0) + \int_{0}^{t}\psi(t-t^{\prime}) P_{B}(x,t,t^{\prime}) \times \nonumber \\ & & \times \left[ \dot{B}(x,t^{\prime}) + j^{-}(x,t^{\prime}) + k (B_0-B(x,t^{\prime})) B(x,t^{\prime}) \right] dt^{\prime} \nonumber \\
P_{B}(x,t,t^{\prime}) &=& \exp\left[ -k \int_{t^{\prime}}^{t}(B_0 - B(x,t^{\prime\prime})) dt^{\prime\prime} \right]. \label{fkpp}
\end {eqnarray}
where the concentration of B is given by $B(x,t)=B_i(t)/a^3$, and the corresponding dimensional
constant is absorbed into a new reaction rate $k=\kappa a^3$. The concentration $B_0$ is set 
to unity in what follows.
Eq.(\ref{fkpp}) can be rewritten in a form following from the
equations of Ref.\cite{vladross} for the irreversible case:
\begin{eqnarray}
\dot B(x,t) &=& - k(1-B(x,t))B(x,t) + \frac{a^2}{2}\Delta \int_0^t M(t-t^\prime) \nonumber \\ && \times B(x,t^\prime) \exp \left[-\int_{t^\prime}^t k(1-B(x,t^{\prime\prime}))dt^{\prime\prime}\right]dt^\prime\label{horstend}
\end{eqnarray}
with $\tilde{M}(u)=u\tilde\psi(u)/[1-\tilde\psi(u)]$.

\subsection{Leading edge linearization}
To analyze the behavior in the leading edge we note that $A(x,t)=1-B(x,t)$ 
becomes small for $x\to\infty$ and hence
\begin{eqnarray}
\frac{\partial B}{\partial t} &=& \frac{a^2}{2}\int_0^t \Delta\left\{ M(t-t^{\prime})B(x,t^{\prime})\exp\left[-k \int_{t^{\prime}}^t (1-B(x,t^{\prime\prime}))\,dt^{\prime\prime}\right] \right\}dt^{\prime} \nonumber \\ 
&& - k(1-B(x,t))\label{basicfkpp}
\end{eqnarray}
where we have interchanged the sequence of differentiation over $x$ and temporal integration.
For $x\to\infty$ we have $\exp\left[-k \int_{t^{\prime}}^t (1-B(x,t^{\prime\prime}))\,dt^{\prime\prime}\right] \rightarrow 1$ so that the integrand can be put into the form
\begin{eqnarray*}
&& M(t-t^{\prime})\Bigg[ \frac{\partial^2 B(x,t^{\prime})}{\partial x^2} + 
2\frac{\partial B(x,t^{\prime})}{\partial x}\int_{t^\prime}^t 
k\frac{\partial B(x,t^{\prime\prime})}{\partial x}\,dt^{\prime\prime} + 
B(x,t^\prime)\int_{t^\prime}^t k\frac{\partial^2 B(x,t^{\prime\prime})}
{\partial x^2}\,dt^{\prime\prime}\\
&& + B(x,t^{\prime})\Big(\int_{t^\prime}^t k\frac{\partial B(x,t^{\prime\prime})}
{\partial x}\,dt^{\prime\prime}\Big)^2 \Bigg].
\end{eqnarray*}

We now assume $B(x,t)$ to be $1-A_0 \exp\left[ -\lambda (x-vt) \right]$ for large $x$, i.e.
consider a wavefront with  an exponentially decaying leading edge
moving at a constant velocity $v$. 
Inserting this into (\ref{basicfkpp}) and retaining only the terms of the 
first order in $A_0$, we get
\begin{eqnarray}
- \frac{\partial}{\partial t} \big(A_0 \exp\left[ -\lambda (x-vt) \right]\big)
&=& -kA_0 \exp\left[ -\lambda (x-vt) \right] \nonumber \\
&& + \frac{a^2}{2}\int_0^t M(t-t^{\prime})\Big[ -\lambda^2 A_0 \exp\left[ -\lambda (x-vt^{\prime}) \right]\nonumber \\ 
&& +\frac{k\lambda}{v} A_0 \exp\left[ -\lambda (x-vt^{\prime}) \right] -\frac{ k\lambda}{v} A_0 \exp\left[ -\lambda (x-vt) \right]  \Big]\, dt^{\prime} \nonumber \\
&=& -kA_0 \exp\left[ -\lambda (x-vt) \right] \nonumber \\
&& + \frac{a^2}{2}A_0\left[-\lambda^2+\frac{k\lambda}{v} \right] \int_0^t M(t-t^{\prime})\exp\left[ -\lambda (x-vt^{\prime}) \right]\, dt^{\prime} \nonumber \\
&& - \frac{a^2}{2}\frac{ k\lambda}{v} A_0 \exp\left[ -\lambda (x-vt) \right] \int_0^t M(t-t^{\prime})\, dt^{\prime},\label{geneq}
\end{eqnarray}
with the memory kernel $\tilde{M}(u)=u\tilde\psi(u)/(1-\tilde\psi(u))$. We note
that the traditional way to proceed is first to linearize the equation and then to
insert the exponential solution, however, since the solution of the linearized equation
is anyhow an exponential we can also proceed the other way around, saving on tedious
calculations.

\paragraph{Markovian case} 

Let us first show that for the Markovian case of exponential waiting time
distribution the standard expression for the minimal velocity of the
stable propagation is reproduced. 
Taking $\psi(t)=\frac{1}{\tau}\exp\left[ -\frac{t}{\tau}\right]$,
one obtains $M(t-t^{\prime})=\frac{1}{\tau}\delta(t-t^{\prime})$ and
\begin{eqnarray} 
\label{cancel}
0 && = \frac{\partial}{\partial t} \big(A_0 \exp\left[ -\lambda (x-vt) \right]\big) 
-kA_0 \exp\left[ -\lambda (x-vt) \right] \nonumber \\
&& + \frac{a^2}{2\tau}A_0 \exp\left[ -\lambda (x-vt) \right] \left[
  -\lambda^2 + \frac{k\lambda}{v} -\frac{k\lambda}{v} \right] \\
&& = \lambda v \exp\left[ -\lambda (x-vt) \right] -kA_0 \exp\left[ -\lambda (x-vt) \right]
- \frac{a^2}{2\tau} \lambda^2 A_0 \exp\left[ -\lambda (x-vt) \right] \nonumber
\end{eqnarray}
which for $z=x-vt$, $z\to\infty$ leads us to the following equation:
\[
\frac{a^2}{2\tau} \lambda^2 -v \lambda + k = 0.
\]
This is the standard dispersion relation for the FKPP front, showing that
real values of $\lambda$ (corresponding to physically sound
solutions with nonnegative concentrations) are only possible for 
$v \geq v_{min} = 2\sqrt{a^2 k / 2\tau} \equiv 2 \sqrt{Dk}$ with
$D=a^2/2\tau$ being the diffusion coefficient (note that $B_0$ is set to unity).
Note that the corresponding result emerges due to the cancellation of
two terms of different nature in Eq.(\ref{cancel}), which, as we proceed to
show, does not take place in the non-Markovian case. 

\paragraph{Non-Markovian case}

We now assume the waiting time pdf to decay as a power law 
$\psi(t)\propto t^{-1-\alpha}$, $0 < \alpha <  1$ for large $t$. 
With $\hat t = t-t^{\prime}$ we get
\begin{eqnarray*}
\int_0^t M(t-t^{\prime})\exp\left[ \lambda vt^{\prime} \right]\, dt^{\prime} &=&
\exp\left[ \lambda vt \right]
\int_0^t M(\hat t)\exp\left[ -\lambda v \hat t \right]\, d\hat t\\
&=& \exp\left[ \lambda vt \right]\tilde{M}(\lambda v).
\end{eqnarray*}
and moreover that 
\begin{eqnarray*}
\int_0^t M(t-t^{\prime})\exp\left[ \lambda vt \right]\, dt^{\prime} &=& \frac{const}{\tau^{\alpha}}t^{\alpha-1},
\end{eqnarray*}
so that the last term in (\ref{geneq}) vanishes for large $t$.
Altogether we have then:
\begin{eqnarray*}
\frac{\partial}{\partial t} \big(1-A_0 \exp\left[ -\lambda (x-vt) \right]\big) &=& -kA_0 \exp\left[ -\lambda (x-vt) \right] \\
&& + \frac{a^2}{2}A_0 \exp\left[ -\lambda (x-vt) \right]\left[ (-\lambda^2 +\frac{ k\lambda}{v})\tilde{M}(\lambda v)\right],
\end{eqnarray*}
and with $z=x-vt$
\begin{eqnarray}
-\lambda v A_0 \exp\left[ -\lambda z \right] &=& -kA_0 \exp\left[ -\lambda z \right] \nonumber \\
&& + \frac{a^2}{2}A_0 \exp\left[ -\lambda z \right]\left[ (-\lambda^2 + \frac{k\lambda}{v})\tilde{M}(\lambda v)\right] ;\nonumber \\
0 &=& - \lambda v + k + \frac{a^2}{2}(\lambda^2 -\frac{ k\lambda}{v})\tilde{M}(\lambda v)\label{nmeq}
\end{eqnarray}
For example, taking $\psi(t)$ to be given by a Mittag-Leffler function 
$\psi(t)=E_\alpha[-(t/\tau)^\alpha]$, we have exactly 
$\tilde\psi(u)=[1+(u\tau)^{\alpha}]^{-1}$ and $\tilde{M}(u)= \tau^{-\alpha}u^{1-\alpha}$, so that
\begin{eqnarray*}
0 &=& - \lambda v +k + \frac{a^2}{2\tau^{\alpha}}(\lambda^2 -\frac{ k\lambda}{v})(\lambda v)^{1-\alpha}.
\end{eqnarray*}
This equation is equivalent to
\begin{eqnarray}
(v\lambda - k)\Big(\frac{a^2}{2\tau^{\alpha}}\lambda^{2-\alpha}v^{-\alpha} -
1\Big) &=& 0
\end{eqnarray}
and always possesses two nonnegative roots. Therefore the minimal propagation
velocity in this case is zero. 
The propagation failure
corresponds essentially to a continuum approaching of $v_{min} \propto \sqrt{D}$
to zero for the case when $D \rightarrow 0$ as it is the case in subdiffusion.  

\section{Comparison with a related model of Ref.\cite{YAD}}

Our result here differs from the one of Yadav et al.\cite{YAD}, where the stable
front propagation with the constant velocity was found. 
The authors of Ref.\cite{YAD} consider a situation in which the A-particles undergo 
a reversible reaction corresponding to the branching-coalescence scheme,
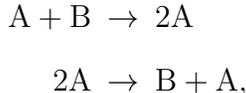
\begin{eqnarray*}
\mathrm{A} + \mathrm{B} &\rightarrow& 2\mathrm{A} \\
2\mathrm{A} &\rightarrow& \mathrm{B} + \mathrm{A},
\end{eqnarray*}
which for the Markovian case is also described by the FKPP equation. 
The equation used in Ref.\cite{YAD} for the non-Markovian case
reads: 
\begin{eqnarray}
\frac{\partial A(x,t)}{\partial t} &=& kA(x,t) - k_1\left[ A(x,t)\right]^2 \nonumber \\
&+& a^2\Delta \left\lbrace \int_0^t M[t-t^{\prime}]A(x,t^\prime) \times 
\exp\left[-\int_{t^{\prime}}^t kA(x,t^{\prime\prime})\,dt^{\prime\prime} \right] 
\,dt^{\prime} \right\rbrace.\label{horstfront}
\end{eqnarray}
The velocity of the front was then obtained by means of the Hamilton-Jacobi approach 
via a hyperbolic scaling (a more elegant method known to lead to the same
results as leading edge linearization in the Markovian case) and reads in our notation
\begin{eqnarray}
v_{min} &=& \frac{k(\alpha-3)}{\alpha-2}\sqrt{\frac{a^2}{\tau}
\left[\frac{k(\alpha-3)}{\alpha-2}\tau\right]^{1-\alpha}\frac{2-\alpha}{k} } = 
\sqrt{k^{2-\alpha}K_{\alpha}\left(\frac{ \alpha-3}
{\alpha-2}\right)^{3-\alpha}(2-\alpha)},
\end{eqnarray}
with a generalized diffusion constant $K_{\alpha}=a^2/\tau^{\alpha}$.

Since the term quadratic in $A$ does not contribute to the linearized
solution determining the pulled front properties, the only difference
between our approach and the one of the authors of
Ref.\cite{YAD} is that Eq.(\ref{horstfront}) is written for the
concentration of the A-particles which are \textit{created} during the
reaction, while our equation Eq.(\ref{horstend}) is put down for the B-particles which 
\textit{irreversibly
disappear}.  Taking Eq.(\ref{horstfront}) for granted, we first checked
that the method of the leading edge linearization used in the previous
sections yields the same result for the minimal front velocity as in
Ref.\cite{YAD}. Therefore the differences between the results are not
connected to the method applied, but to the equation itself.

The difference in the results is explained by the fact that the
equations for particles which are created during the reaction under
the conditions discussed look differently from Eq.(\ref{horstfront}):
Since the particle making its step as A might have entered the
corresponding site both as A and as B (and having undergone a transformation),
the equation for the product of the reaction (in our case A) has to
contain the Laplacians of both $A$ and $B$-concentrations, as
discussed in \cite{SSS}.  Moreover, considering reversible
reactions put additional problems due to possible multiple $\mathrm{A}
\rightleftharpoons \mathrm{B}$ transformations during the waiting time
at one site \cite{sashso}. Although leading to the classical FKPP-results in the
Markovian case, the equations proposed in
\cite{vladross} are not well suited for the description of particles
created during the reaction and for reversible reactions under
conditions discussed in the beginning of the paper.

\section{Conclusion}

Let us summarize our findings. We considered the reaction-subdiffusion
problem for the irreversible autocatalytic conversion reaction
$\mathrm{A}+ \mathrm{B} \rightarrow 2 \mathrm{A}$ which, in the case
of normal diffusion, is described by the FKPP equation. We show that
in the case of subdiffusion, under the assumption of the reaction
taking place irrespectively on the particles' mesoscopic motion, 
the minimal front's propagation velocity is zero, suggesting the propagation
failure. This regime can be considered as a result of a
continuous transition to $D \rightarrow 0$ in the normal FKPP situation.
The result obtained is in contrast to the treatment in \cite{YAD}
where the finite propagation velocity of a traveling wave was found.
The differences in the assumptions of the two approaches (corresponding
essentially to two \textit{very different situations}) are discussed.

\section{Acknowledgments}

D. Froemberg and I.M. Sokolov gratefully acknowledge the financial support
of DFG within the SFB555 research collaboration.  F. Sagu\'es
acknowledges financial support from MEC under project FIS 2006 -
03525 and from DURSI under project 2005 SGR 00507.

\end{document}